\documentstyle[12pt]{article}
\advance\textheight by 60pt
\advance\voffset by -40pt
\advance\textwidth by 50pt
\advance\oddsidemargin by -25pt
\advance\evensidemargin by -25pt

\def\single_space{\baselineskip 12pt plus 1pt minus 1pt}
\def\one_and_a_half_space{\baselineskip 19pt plus 1pt minus 1pt}
\def\double_spacesp{\baselineskip 25pt plus 2pt minus 2pt}

\def\atversim#1#2{\lower0.7ex\vbox{\baselineskip\zatskip
\lineskip\zatskip  \lineskiplimit 0pt\ialign{$\matth#1
\hfil##\hfil$\crcr#2\crcr\sim\crcr}}}

\begin{document}
\begin{titlepage}
\begin{flushright}
{\bf
ANL-HEP-PR-94-58\\
November  1994\\
}
\end{flushright}
\vskip 1.5cm
{\Large
{\bf
\begin{center}
Aspects of four-jet production \\
in  polarized proton-proton collisions
\end{center}
}}
\vskip 1.0cm
\begin{center}
S. P. Fraser and S. T. Fraser\footnote{Address as of Sept. 1, 1994,
Univ. Heidelberg, D-W-6900,
Heidelberg 1 GERMANY}
\\
Department of Physics and Astronomy\\
Sonoma State University \\
Rohnert Park, CA 94928 USA \\
\vskip .3cm
and\\
\vskip .3cm
R. W. Robinett \\
High Energy Physics Division \\
Argonne National Laboratory \\
Argonne, IL 60439  USA\\
\vskip 0.05cm
and
\vskip 0.05cm
Department of Physics \\
Penn State University\\
University Park, PA 16802 USA \\
e-mail:rq9@psuvm.psu.edu\\
\end{center}
\vskip 1.0cm
\begin{abstract}

We examine the intrinsic spin-dependence of the dominant
$gg \rightarrow gggg$
subprocess contribution to four-jet production in polarized proton-proton
collisions using helicity amplitude techniques.
 We find that the partonic level, longitudinal spin-spin
asymmetry, $\hat{a}_{LL}$, is intrinsically large in the kinematic
regions probed in experiments detecting four isolated jets.
Such events may provide another qualitative or semi-quantitative
test of the spin-structure of QCD in planned polarized $pp$ collisions
at RHIC.

\end{abstract}
\end{titlepage}
\double_spacesp

The prospects for a comprehensive program of polarized proton-proton
collisions at collider energies at RHIC \cite{particleworld,workshop},
culminating in the
approved experiment R5 \cite{r5}, has motivated a large number of
studies of the spin-dependence of many standard model processes and their
sensitivity to polarized parton distributions. Many processes
which already have been well-studied, both
theoretically and experimentally, at existing unpolarized hadron colliders,
have been reexamined in the context of a physics program dedicated to the
extraction of the spin-dependent quark, antiquark, and gluon distributions
and tests of the spin-dependence of the basic hard scattering processes
of QCD and the electroweak sector.

Familiar processes such as direct photon production
\cite{photons,soffer1991,zphys} and Drell-Yan
lepton pair production \cite{zphys,drell-yan}
(including $W$ and $Z$ production \cite{zphys,soffer1994})
are known to be sensitive to the longitudinally polarized
gluon and sea-quark content of the proton respectively.
(We will henceforth take polarized to mean longitudinal polarization;
transverse polarization effects will also be studied at RHIC but will
only be briefly discussed here.)
Calculations of the radiative
corrections to the spin-dependent cross-sections for these processes
have even appeared \cite{radiative1,radiative2} and confirm the general
conclusions of leading-order results.
Jet production has also been extensively studied
at lowest order \cite{soffer1991,jets} and the technology for the
efficient
calculation of the NLO corrections to helicity amplitudes for
jet production is now well-known \cite{bern}
although a detailed analysis of the spin-dependent
radiative corrections to jet cross-sections has not yet been
performed.  Taken together, these processes already
provide the basis for a substantial experimental program using the
two planned RHIC detectors \cite{r5},
STAR and PHENIX, which are complementary
in their physics capabilities.

A wide variety of other processes has also
been studied at leading-order
in the context of
the RHIC spin program including heavy quark production
\cite{contogouris,karliner},
           quarkonium production in different kinematic regions
\cite{cortes,rwr1,rwr2}, three-jet production \cite{rwr3}, and
double-photon ($\gamma \gamma$) production \cite{rwr4}.  All of these
processes have been studied experimentally, to varying degrees,
in collider energy  unpolarized $pp$ or $p\overline{p}$ collisions and
seem to agree reasonably well with theoretical expectations.
Some other processes whose spin-dependence has also been discussed,
such as
two-jet plus direct photon \cite{rwr3} and $\psi + \gamma$
\cite{dkim}, have yet to be measured but may be detectable for the
first time given the
large luminosity and energies (up to $2\times10^{32}\,cm^{-2}s^{-1}$
and $\sqrt{s} = 500\,GeV$) possible at RHIC and the nature of the
proposed detectors.

Just as with single inclusive jet, di-jet, and three-jet production,
four-jet production has been observed at hadron colliders ranging from
the ISR \cite{breakstone,afs} ($\sqrt{s} = 63\,GeV$) to the
CERN $Sp\overline{p}S$ \cite{ua2} ($\sqrt{s} = 630\,GeV$)
and most recently at the TEVATRON \cite{cdf} ($\sqrt{s}
= 1.8\,TeV$).  In the last two cases,
comparisons of data to leading order QCD predictions for
various shape and angular variables
have been made and reasonable agreement
is found.  Besides providing another test of standard QCD, this
process has the additional feature that it
has also been widely discussed as a possible arena
in which to study double-parton scattering \cite{double,mangano}, an
effect which is expected to appear at sufficiently high energies.

Because the $2 \rightarrow 4$ subprocesses leading to four-jet production
are proportional to $\alpha_S^4$, changes
in the value of $\alpha_S$ used and
uncertainties in the choice of $Q^2$ scale
(i.e. lack of knowledge of the next order QCD corrections) make the
prediction of the absolute rates difficult.
  This coupled with the lower rates
compared to two- and three-jet production and the increasing difficulty
of unambiguously defining $n$ isolated jets makes this process, at present,
a semi-quantitative or qualitative test of QCD.
Because rapid progress
is being made in the calculation of the NLO matrix elements for two- and
three-jet
production \cite{bern} with hope held out for the eventual evaluation of
the necessary radiative corrections for the $2 \rightarrow 4$ subprocesses
as well \cite{dixon}, this situation may well change
in the not-too-distant future.

Motivated by the large rates possible for many hadron processes at the
high luminosity polarized $pp$ RHIC facility, in this note we briefly
discuss the spin-dependence of the dominant QCD subprocess contributing
to such four-jet production, namely $gg \rightarrow gggg$; the analysis
presented here thus extends and complements
the discussion of the spin-dependence of three-jet production given
in Ref.~\cite{rwr3}.  Because existing limits \cite{ua2,cdf} imply
that double-parton scattering will not be important at any RHIC
energy, we will discuss only the spin-dependence of $2 \rightarrow 4$
processes.

We follow the analysis of four-jet production at TEVATRON
energies as  given in
Ref.~\cite{mangano} with some minor differences.  We consider
only the dominant $gg \rightarrow gggg$ subprocess and because we
are interested in the spin-dependence we use the exact expressions
for the matrix elements for this process instead of relying on any
approximation technique; specifically, we use the
compact expressions for the necessary helicity amplitudes
in Ref.~\cite{kuijf} and Ref.~\cite{berends}.
In the calculation of the appropriate cross-sections,
we modify the cuts of Ref.~\cite{mangano} Set 2 (motivated by
the smaller RHIC energy of $\sqrt{s} = 500\,GeV$)  and insist that
$p_T > 15\,GeV$ and $|\eta| \leq 0.8$ for each jet and that
$\cos(\theta_{ij}) \leq 0.643$ for each jet pair.  Finally, we insist
that there is a minimum total transverse energy of $E_T \geq 70\,GeV$.
For this leading-order calculation, we use the updated parton
distributions of Duke-Owens
(in this case, only the gluon distribution of Set I is required) of
Ref.~\cite{dukeowens}.

The resulting differential cross-section as a function of transverse
momentum, $d\sigma /dp_T$ versus $p_T$, (with four entries per event)
is shown as the dot-dash curve in Fig.~1 and has a very similar shape
as the corresponding plot (Fig.~2~(a))
in Ref.~\cite{mangano}. The spin-dependent
cross-sections
\begin{equation}
\frac{d \Delta \sigma}{dp_T} \equiv \frac{1}{2}
\left( \frac{d \sigma (++)}{dp_T} - \frac{d \sigma (+-)}{dp_T}
\right)
\end{equation}
where $(++),\,(+-)$ refers to the helicities of the incident protons
are also shown in Fig.~1 for two choices of the polarized gluon
distribution.
  The spin-dependent gluon distribution,
conventionally written as
$\Delta G(x,Q^2) \equiv G_{+}(x,Q^2) - G_{-}(x,Q^2)$
(where $+/-$ refers to parton helicity in the same/opposite direction
as the parent proton helicity), is assumed, for simplicity, to have
the form $\Delta G(x,Q^2) = x^{\alpha} G(x,Q^2)$.  The two choices
$\alpha = 1$ (solid curve) and $\alpha = 0.25$ (dashed curve) correspond
to an integrated gluon contribution $\Delta G = \int \Delta G(x,Q^2)\,dx$
equal to $0.5$ and $4.5$ respectively and thus bracket the expectations
for a 'normal' to 'large' gluonic contribution to the total proton spin.

Using the $gg \rightarrow gggg$ matrix elements and these cuts, we have
calculated
differential distributions in the other kinematic variables discussed in
Ref.~\cite{mangano}, namely $p_{out}$, $\phi_{min}$, and
$\cos(\theta_{23}^*)$. The transverse momentum out of the plane
passing through the beam and the jet of largest $p_T$, i.e.
$p_{out}$, is defined via
\begin{equation}
p_{out} \equiv \frac{1}{2} \sum_i|p_{out}^{i}|
\end{equation}
while
$\phi_{min}$ is the minimum angle in the transverse plane
between the largest $p_T$ jet and the other three jets and finally
$\cos(\theta_{23}^*)$ is the cosine of the angle between the second
and third most energetic jets in the
four-jet center-of-mass.  Differential
distributions in each of these quantities look very similar to the
corresponding plots in Fig.~2(a)-(d) in Ref.~\cite{mangano}.

The measurable spin-spin asymmetry
in any observable quantity, defined by
\begin{equation}
A_{LL} =
\frac{d\sigma(++) - d\sigma(+-)}{d\sigma(++) + d\sigma(+-)}
\end{equation}
is determined by both the partonic level spin-dependence of the
underlying hard scattering,
\begin{equation}
\hat{a}_{LL} = \frac{d\hat{\sigma}(++) - d\hat{\sigma}(+-)}
{d\hat{\sigma}(++) + d\hat{\sigma}(+-)}
\end{equation}
and the magnitude of the polarized parton distributions
since
\begin{equation}
A_{LL} d \sigma = \sum_{i,j} \int dx_a \int dx_b \;d \hat{\sigma}
\;\hat{a}_{LL}\;
\Delta f_i(x_a,Q^2) \;\Delta f_j(x_b,Q^2)
\end{equation}
Using the polarized parton distributions mentioned above and the
exact $2g \rightarrow 4g$ matrix elements, we find the observable
asymmetries for the four variables $p_T$, $p_{out}$, $\phi_{min}$,
and $\cos(\theta_{23}^*)$ shown
in Fig.~2(a)-(d).  The increasing value
of $A_{LL}$ with larger values of $p_T$ (Fig.~2(a)) and
$p_{out}$ (Fig.~2(b)) is reminiscent
of similar effects in two-jet \cite{soffer1991,jets} and
three-jet \cite{rwr3} production and simply reflects the increase
in gluon polarization for large
$x$ (where, of course, the cross-sections
are very small).  Figs.~2(a) and (b) also
show that the intrinsic spin-spin
asymmetry of the hard-scattering subprocess, $\hat{a}_{LL}$,
 is large; we estimate that
the average value of the partonic level spin-spin asymmetry in the
configurations measured in four-jet production is roughly
$\langle \hat{a}_{LL} \rangle
\sim 0.8$ which is even larger then  the
corresponding value of $\langle \hat{a}_{LL} \rangle \sim 0.7$
found for the $gg \rightarrow ggg$ contribution to three-jet
production.
We have also checked explicitly that when one gluon is
allowed to be soft the resulting partonic level spin-spin
asymmetries reduce to those for three-jet production derived in
Ref.~\cite{rwr3}.
On the other hand, the angular variables we
have studied show little variation
for a given
set of polarized distribution functions
as seen in Fig.~2 (c) and (d); this  is also consistent with
earlier results on spin-dependence in three-jet events.
These results reaffirm the fact observed in
the case of two- and three-jet production, namely
that once one is in the kinematic
regime required for the isolation of jets, the intrinsic longitudinal
spin-spin asymmetries for gluon initiated processes is almost maximally
large and varies little with the relative orientation of the jets.

One can also consider transverse spin effects in jet physics but the
situation is far less diverse.  The quarks and antiquarks (and not the
gluons) carry
the  'transversity' \cite{jaffe-ji} so that only  $qq$ or $q\overline{q}$
initiated processes in transversely
polarized $pp$ collisions
need be considered.  It has been known for some time \cite{sivers}
that the
partonic level transverse spin-spin asymmetries, $\hat{a}_{TT}$,
for the relevant $2 \rightarrow 2$ processes vanish in the
case of unlike quark scattering via $q q' \rightarrow q q'$ and are
much smaller (roughly a factor of $10$)
than the corresponding longitudinal asymmetry ($\hat{a}_{LL}$)
for like-quark scattering via $q q \rightarrow q q$ due to a color factor.
An identical  pattern is seen in the transverse spin-spin asymmetries for
three-jet production via $q q' \rightarrow q q' g$ and
$q q \rightarrow q q g$ \cite{rwr5}.  We expect a similar situation
in the four-jet case for the processes $q q' \rightarrow q q' g g$,
$q q \rightarrow q q g g$ and $q q' \rightarrow q q' q q$.  Since these
processes only form a significant part of the four-jet cross-section
at the very largest values of $p_{T}$ where the cross-sections are
unmeasureably small, we expect no significant transverse spin dependence
in such processes.

\vskip 0.5cm

\begin{flushleft}
{\large {\bf Acknowledgments}}
\end{flushleft}

One of us (RWR) gratefully acknowledges the
hospitality of Argonne National Laboratory where this work was completed.
This work was supported, in part, by a grant from the National Science
Foundation under the Research Experiences for Undergraduates (REU)
program (SPF, STF) at Penn State,
by Argonne National Laboratory under the Faculty
Research Participation Program (RWR), and by the U. S. Department of
Energy, Division of High Energy Physics, Contract W-31-109-ENG-38 (RWR).

\newpage
\single_space
\def\etal{{\em et al.}}
\def\PL{{ Phys. Lett.\ }}
\def\NP{{ Nucl. Phys.\ }}
\def\PR{{ Phys. Rev.\ }}
\def\PRL{{ Phys. Rev. Lett.\ }}

\newpage
\double_spacesp

{\Large Figure Captions}

\begin{itemize}

\item[Fig. \thinspace 1.] The solid and dashed curves show the
spin-dependent, differential cross-section,
$d\Delta \sigma/ dp_T (nb/GeV)$ versus $p_T (GeV)$,
for four-jet production from
the $gg \rightarrow gggg$ subprocess
with the cuts discussed in the text; the solid (dashed) curves
correspond to a polarized gluon distribution given by
$\Delta G(x,Q^2) = x^{\alpha} G(x,Q^2)$ with $\alpha = 1\,(0.25)$.
The dot-dash curve is the unpolarized differential cross-section
$d\sigma/ dp_T (nb/GeV)$ versus $p_T (GeV)$ with the same assumptions.

\item[Fig. \thinspace 2.] The spin-spin asymmetry,
$A_{LL}$, in various
differential cross-sections for four-jet production.  The plots
correspond to asymmetries in
differential distributions  for   (a) $p_T$, (b) $p_{out}$,
(c) $\phi_{min}$, and (d) $\cos(\theta^*_{23})$.
The solid (dashed) curves
in each plot correspond to polarized gluon distributions given by
$\Delta G(x,Q^2) = x^{\alpha} G(x,Q^2)$ with $\alpha = 1\,(0.25)$.

\end{itemize}
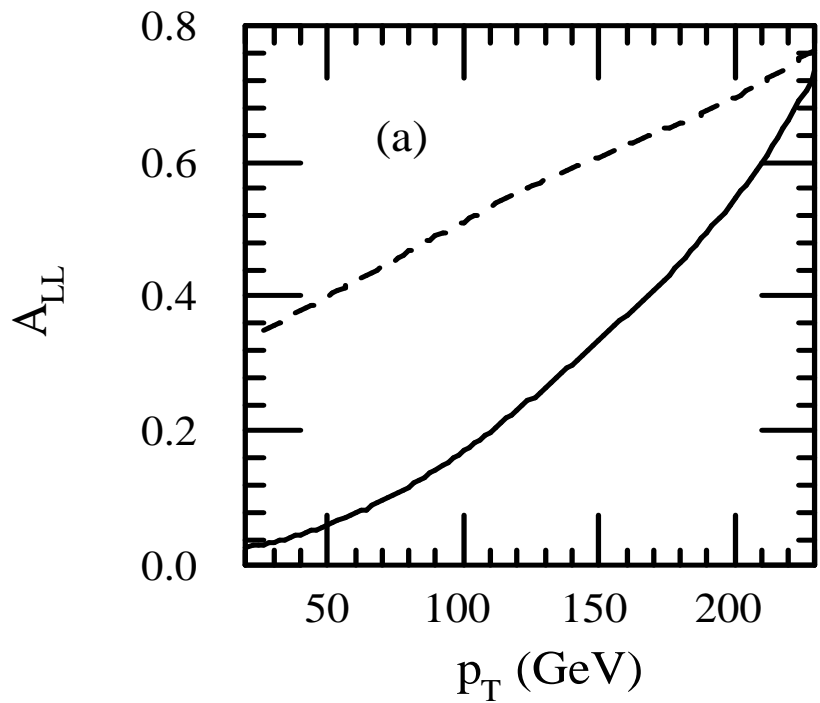
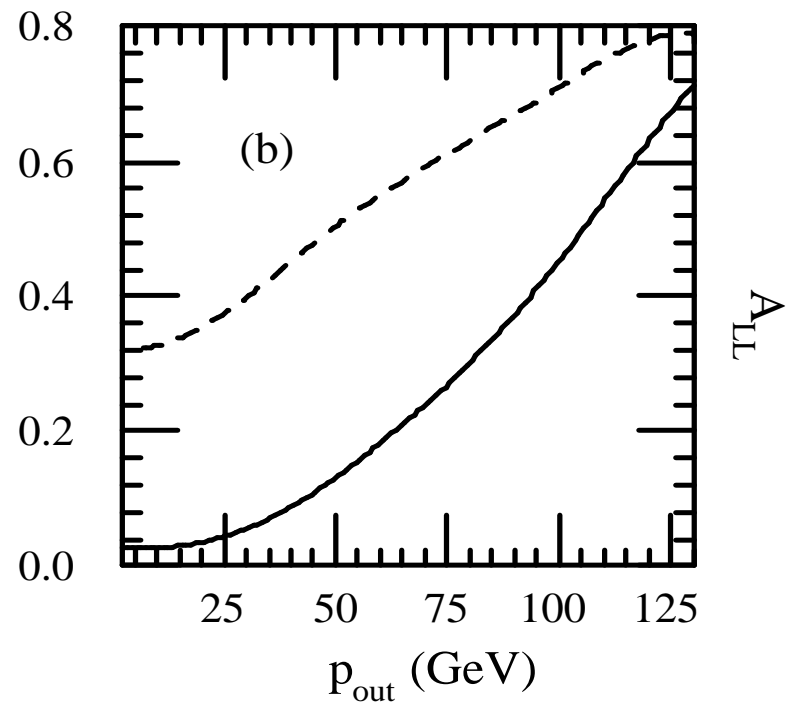
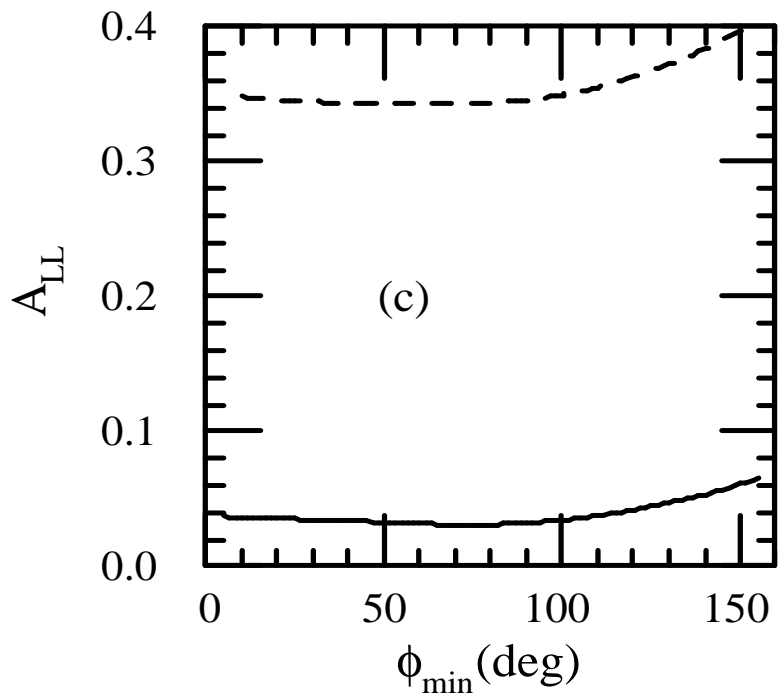
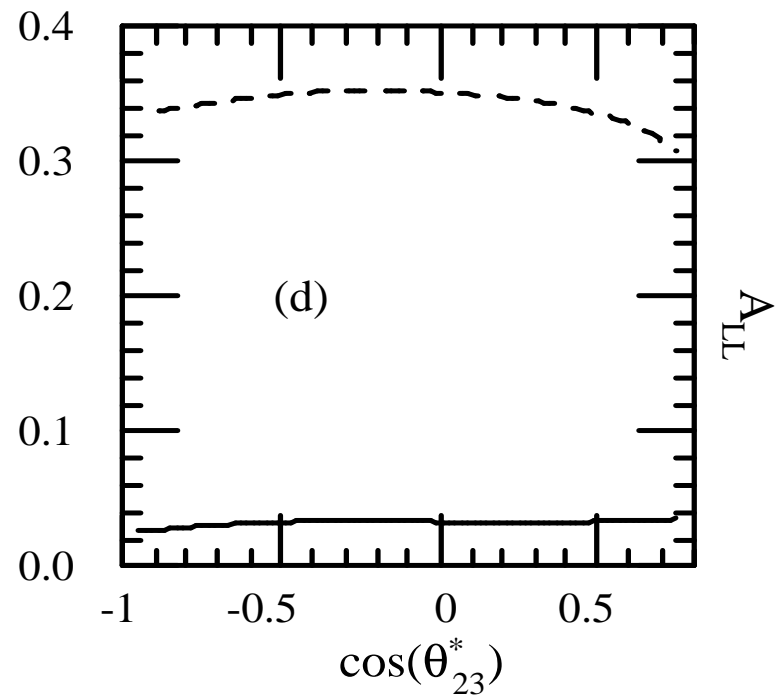
\end{document}